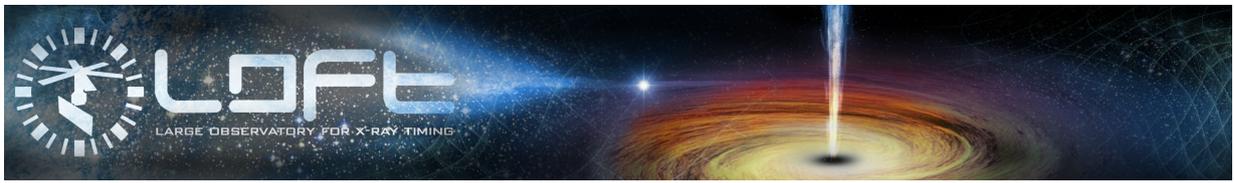

# The study of neutron star magnetospheres with *LOFT*

White Paper in Support of the Mission Concept of the Large Observatory for X-ray Timing

Authors

R.P. Mignani[1,2], F. Bocchino[3], N. Bucciantini[4], M. Burgay[5], G. Cusumano[6],
A. De Luca[1], P. Esposito[1,7], C. Gouiffes[8], W. Hermsen[9], G. Kanbach[10],
L. Kuiper[9], G.L. Israel[11], M. Marelli[1], S. Mereghetti[1], T. Mineo[6], C. Motch[12],
A. Pellizzoni[5], A. Possenti[5], P.S. Ray[13], N. Rea[14,15], B. Rudak[16], D. Salvetti[1],
A. Shearer[17], A. Słowikowska[2], A. Tiengo[18,19], R. Turolla[20,21], N. Webb[22]

[1] INAF, IASF Milan, Milano Italy
[2] University of Zielona Gora, Poland
[3] INAF, OA Palermo, Palermo, Italy
[4] INAF, OA Arcetri, Italy
[5] INAF, OA Cagliari, Cagliari, Italy
[6] INAF, IASF Palermo, Palermo, Italy
[7] Harvard-Smithsonian Center for Astrophysics, Cambridge, MA, USA
[8] CEA, Saclay, France
[9] SRON, Utrecht, Netherlands
[10] Max Planck Institut für extraterrestrische Physik, Garching bei München, Germany
[11] INAF, OA Rome, Rome, Italy
[12] Observatoire Astronomique de Strasbourg, Strasbourg, France
[13] Navy Research Laboratory, Washington, DC, USA
[14] University of Amsterdam, Amsterdam, Netherlands
[15] IEEC-CSIC, Barcelona, Spain
[16] Copernicus Astronomical Center, Warsaw, Poland
[17] National University of Ireland, Galway, Ireland
[18] IUSS Pavia, Pavia, Italy
[19] INFN, Pavia, Italy
[20] University of Padova, Padova, Italy
[21] MSSL-UCL, London, UK
[22] IRAP, Toulouse, France

**Preamble**

The Large Observatory for X-ray Timing, *LOFT*, is designed to perform fast X-ray timing and spectroscopy with uniquely large throughput (Feroci et al. 2014). *LOFT* focuses on two fundamental questions of ESA's Cosmic Vision Theme "Matter under extreme conditions": what is the equation of state of ultra-dense matter in neutron stars? Does matter orbiting close to the event horizon follow the predictions of general relativity? These goals are elaborated in the mission Yellow Book (http://sci.esa.int/loft/53447-loft-yellow-book/) describing the *LOFT* mission as proposed in M3, which closely resembles the *LOFT* mission now being proposed for M4.

The extensive assessment study of *LOFT* as ESA's M3 mission candidate demonstrates the high level of maturity and the technical feasibility of the mission, as well as the scientific importance of its unique core science goals. For this reason, the *LOFT* development has been continued, aiming at the new M4 launch opportunity, for which the M3 science goals have been confirmed. The unprecedentedly large effective area, large grasp, and spectroscopic capabilities of *LOFT*'s instruments make the mission capable of state-of-the-art science not only for its core science case, but also for many other open questions in astrophysics.

*LOFT*'s primary instrument is the Large Area Detector (LAD), a $8.5\,\mathrm{m}^2$ instrument operating in the 2–30 keV energy range, which will revolutionise studies of Galactic and extragalactic X-ray sources down to their fundamental time scales. The mission also features a Wide Field Monitor (WFM), which in the 2–50 keV range simultaneously observes more than a third of the sky at any time, detecting objects down to mCrab fluxes and providing data with excellent timing and spectral resolution. Additionally, the mission is equipped with an on-board alert system for the detection and rapid broadcasting to the ground of celestial bright and fast outbursts of X-rays (particularly, Gamma-ray Bursts).

This paper is one of twelve White Papers that illustrate the unique potential of *LOFT* as an X-ray observatory in a variety of astrophysical fields in addition to the core science.

# 1 Summary

*LOFT* (Large Observatory For X-ray Timing) is a candidate ESA Medium-size mission. In this White Paper we discuss its role in the study of isolated neutron stars, as part of the mission Observatory Science program. *LOFT* will uniquely conjugate the diagnostic power of very high X-ray time and spectral resolution to:

- precisely map the topology of the radiation emitting regions in neutron star magnetospheres, from the analysis of the X-ray light curves;

- directly measure the intensity of neutron star magnetic fields and their variations as the star rotates, from the detection of cyclotron spectral lines;

- investigate variations in the neutron star magnetospheres down to sub-millisecond time scales, tracked by sudden changes in the X-ray light curves.

This is key to understand both the structure and stability of neutron star magnetospheres and the complex radiation emission processes in magnetic field regimes unattainable on Earth - a problem that has challenged astronomers since the discovery of neutron stars. Furthermore, *LOFT* will also:

- discover new, possibly radio-quiet, isolated neutron stars in unidentified *Fermi* and Cherenkov Telescope Array gamma-ray sources, from their pulsed X-ray emission;

- discover peculiar isolated neutron stars, like the magnetars, from their transient and bursting X-ray emission, and monitor their flux, spectral, and pulse evolution;

- work in synergy with the next generation of large radio telescopes, e.g. the Square Kilometre Array, in the X-ray follow-up of newly-discovered radio pulsars.

This wealth of discoveries will, on one hand, unveil the nature of unidentified Galactic gamma-ray sources, on the other, will improve neutron star population synthesis models, mostly verified against radio-selected object samples.

# 2 Introduction

The Large Observatory For X-ray Timing (*LOFT*) has been a contender in the last ESA Medium-size mission selection (M3). The main *LOFT* characteristics are summarised by Feroci et al. (2014) and its two instruments, the Large Area Detector (LAD) and the Wide Field Monitor (WFM), are described by Zane et al. (2014) and Brandt et al. (2014), respectively. A slightly updated configuration of *LOFT* is being proposed for the current ESA selection (M4). Additional and updated information on the *LOFT* M4 configuration can be found in the *LOFT* Consortium web site (http://www.isdc.unige.ch/loft). As part of the Mission Observatory Science Programme, *LOFT* will observe different classes of X-ray sources. In this White Paper, we describe its contribution to the study of isolated neutron stars (INSs; with this term we refer to non-accreting neutron stars, be they solitary or in binary system).

INSs form when the cores of massive stars ($> 10\,M_\odot$) collapse to nuclear densities, acquiring magnetic fields thousand billion times stronger than our Sun. According to the magnetic dipole model, their fast rotation ($\sim 1$ ms–10 s periods) powers the acceleration of particles in the magnetosphere and the emission of electromagnetic radiation beamed around the magnetic axis (e.g., Curtis Michel 1991 and references therein), producing the characteristic pulsed radio emission (hence the name pulsar) if the magnetic and rotation axis are misaligned (see, Lorimer & Kramer 2012 for a review on radio pulsars). Thus, INSs are ideal laboratories to test theories of nuclear physics and electromagnetism under extreme conditions. Almost 50 years after the discovery of the first radio pulsar (Hewish et al. 1968), nearly 2500 INSs are known (http://www.atnf.csiro.au/people/pulsar/psrcat/, see Manchester et al. 2005), both in the Galaxy and in the Magellanic Clouds. However, it is now clear that rotation-powered persistent radio pulsars are not representative of the entire INS population and that their phenomenology is more complex than initially thought. First of all, not all radio pulsars are persistent sources: in some the radio emission is intermittent, in others it is erratic or transient, as in the so-called Rotating Radio Transients (RRaTs). Secondly, pulsars are not just radio sources but are also observed in gamma rays, X rays, and in the optical. Thirdly, some pulsars are not even detected as radio sources, hence dubbed radio quiet (RQ), and are discovered thanks to their high-energy emission. Fourthly, some INSs are powered by an energy source other than the neutron star rotation, e.g., the magnetic energy in the case of the magnetars. Last but not least, in some RQ INSs the emitted radiation does not even come from the magnetosphere but from the hot stellar surface. Based on their characteristics, INSs are classified in five main families, recognised by more or less different locations in the spin period-period derivative diagram (Fig. 1). Their differences and multi-wavelength phenomenology are summarised in Harding (2013), whereas specific reviews are found in Becker et al. (2009).

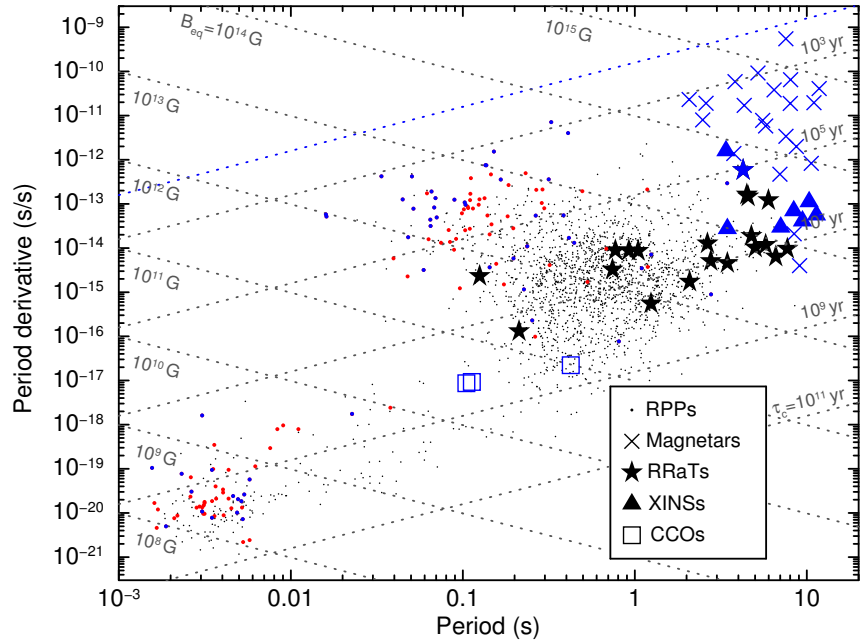

Figure 1: Spin period-period derivative ($P$–$\dot{P}$) diagram for INSs. Different INS families are marked with different symbols: rotation-powered pulsars (RPPs), magnetars, INSs with only thermal X-ray emission (XINSs), Rotating Radio Transients (RRaTs), Central Compact Objects (CCOs) – see Harding (2013). INSs detected in X rays are marked in red, those also showing pulsed X-ray emission are marked in blue. Lines of constant characteristic age ($\tau_c = P/(2\dot{P})$) and equatorial magnetic field ($B_{eq} = 3.2 \times 10^{19}(P\dot{P})^{1/2}$ G), inferred from the magnetic dipole model (e.g., Curtis Michel 1991 and references therein), are indicated by the dotted lines. Since both the LAD and WFM are sensitive above 2 keV, all INSs, except the XINSs and CCOs with very soft X-ray spectra, will be studied with LOFT.

Since several INSs are X-ray sources (Fig. 1), they are obvious targets for *LOFT*. To exploit its capabilities at best, we focus on INSs with strong magnetosphere emission and harder X-ray spectra, i.e., rotation-powered pulsars (hereafter, RPPs) and magnetars. Thanks to its sensitivity over a wide energy range (2–50 keV), large collecting area (8.5 m$^2$ at 8 keV), high time (10 $\mu$s) and spectral resolution (260 eV at 6 keV), the LAD will make it possible to fully exploit the diagnostic power of high X-ray time and spectral resolution to determine the characteristics of neutron star magnetic fields, the geometry of their magnetospheres, study the radiation emission processes in different regimes of magnetic field strengths, and investigate variability phenomena down to millisecond time scales, related to either reconfiguration of the magnetic field or changes in the magnetosphere properties. Furthermore, *LOFT* will play a key role in the discovery of new RPPs and magnetars. The LAD will search for X-ray pulsations from unidentified RPPs associated with gamma-ray sources detected, e.g., by *Fermi* and the Cherenkov Telescope Array (*CTA*), and from RPPs discovered in radio by, e.g., the Square Kilometer Array (*SKA*). In parallel, the WFM will scan the sky in search of new magnetars, or yet unknown types of variable INSs, discovered from the detection of X-ray bursts or outbursts.

In this White Paper we describe how *LOFT* will contribute to the study of the persistent and erratic emission in RPPs (Sects. 3 and 4), the transient emission in magnetars (Sect. 5), and the discovery of new RPPs and magnetars (Sect. 6). All simulations presented here are based on the most recent LAD and WFM response and background files from the *LOFT* Consortium.

## 3 Steady emission in the pulsar magnetosphere

Understanding how the complex radiation processes in pulsar magnetospheres work – a problem that has eluded astronomers for almost 50 years – is one of the major goals in neutron star astrophysics. In particular, it is not completely established yet whether (and how) the emission at different energies (gamma rays, X-rays, optical, radio) originates from different populations of relativistic particles and different regions (and altitudes) of the neutron star magnetosphere, as suggested by, e.g., the Outer Gap (Romani 1996) or the Slot Gap (Muslimov & Harding 2004) models, nor what causes the presence of breaks in the pulsar multi-wavelength power-law (PL) spectra. The light curve characteristics at different energies, such as the number of peaks, their widths and morphology, phase separation, and relative phase lags, are very diverse. Such differences encode information on the orientation of the different emission beams with respect to the neutron star spin axis. Thus, from the comparison of the pulsar light curves at different energies one can map the location of different emission regions in the neutron star magnetosphere, hence explain the diversity between RQ and radio loud (RL) pulsars, track the particle energy and density distribution in the magnetosphere, and measure the neutron star spin axis/magnetic field orientation. Therefore, a precise characterisation of pulsar light curves in different energy ranges is key to build self-consistent emission models of the neutron star magnetosphere (see Pierbattista et al. 2014 and references therein).

Over 160 gamma-ray pulsars (https://confluence.slac.stanford.edu/display/GLAMCOG/Public+List+of+LAT-Detected+Gamma-Ray+Pulsars) have been identified by the Large Area Telescope (LAT) aboard *Fermi*, and 25% of them are RQ (see Caraveo 2014 for a recent review). Thus, gamma-ray detections of RPPs outnumber those in any energy range other than radio. With the *Fermi* mission extended up to 2018, this rich harvest is set to continue and provide us with a uniquely large and diverse sample to characterise the pulsar light curves from the gamma rays to the optical. In the X-rays, only 22 young/middle-aged *Fermi* pulsars show pulsed emission, whereas another 36 have only been detected (Marelli et al. 2014). For very few of these pulsars can better X-ray light curves be measured and X-ray pulsations discovered with *XMM-Newton* and *Chandra*, and by investing ∼100–200 ks observing time per target. The LAD will accurately measure X-ray light curves and detect pulsations for many more *Fermi* pulsars and much more efficiently. In the optical, only five of the *Fermi* pulsars are seen to pulsate and two more have just been detected (Mignani 2010). By the time *LOFT* will fly, however, the future generation of large optical telescopes, such as the 39 m$^2$ European Extremely Large Telescope (E-ELT), and the development of new detectors for high time resolution optical photometry will help filling the present gap with respect to other energies.

We simulated the sensitivity to the detection of X-ray pulsations at the $\gamma$-ray pulse period with the LAD in the 2–10 keV energy band assuming a single-peak X-ray light curve with a Lorentzian profile, variable pulsed fraction (PF), and peak full width half maximum (FWHM). We assumed a pulsar PL X-ray spectrum with an average photon index $\Gamma_X = 1.5$ and conservatively kept the hydrogen column density $N_H$ fixed to the Galactic value along the line of sight. For instance, in case of a 50% PF and 0.1 FWHM, in 10 ks we can detect X-ray pulsations, compute the pulsed fraction (10% error), and the peak phase (0.02 error) for pulsars with unabsorbed X-ray flux as low as $3 \times 10^{-13}$ erg cm$^{-2}$ s$^{-1}$ (0.5–10 keV). In case of a sinusoidal profile, in 10 ks we can reach the same accuracy in the light curve characterisation for pulsars with an unabsorbed X-ray flux as low as $6 \times 10^{-13}$ erg cm$^{-2}$ s$^{-1}$ (0.5–10 keV). This means that we can obtain X-ray light curves with the accuracy claimed above for about 40 of the young/middle-aged *Fermi* pulsars currently detected in the X rays. This number increases to about 70 when including old, recycled milli-second pulsars (MSPs; Fig. 1, lower left). For about 50 of these pulsars the sensitivity to the detection of X-ray pulsations extends to the whole LAD energy range (2–50 keV), which makes it possible to increase by a factor of five the number of RPPs seen to pulsate also in the hard X rays.

## 4 Erratic and bursting emission phenomena in pulsar magnetospheres

Whereas most RPPs are observed as persistent emitters, others show erratic emission phenomena on a variety of time scales. Many of such phenomena have been observed in radio pulsars, such as giant pulses, nulling, intermittence, mode changes, and bursts (see, e.g., Keane 2013 for a recent review). Searching for correlation between erratic variations in the peak intensity of single radio pulses and a similar variability at other energies is crucial for constraining the stability of particle acceleration mechanisms in the pulsar magnetosphere and tracking fast variations in its configuration.

Giant Radio Pulses (GRPs) occur in a handful of radio pulsars, mostly young RPPs and, more rarely, old, recycled MSPs. They can be defined as single pulse emission with an intensity that is significantly higher than the average one. Early observations of GRPs indicated a possible correlation between GRP phenomena and the magnetic field strength at the light cylinder, whose radius is the distance of the last closed magnetic field line, possibly pointing to an outer magnetosphere origin. A correlation between Giant Pulses (GPs) in the radio and at higher energies has only been observed in the optical for the Crab pulsar (Shearer et al. 2003; Collins et al. 2012; Strader et al. 2013), where optical GPs show a ~3% increase relative to the average peak flux intensity. GPs have not been detected yet in the high-energy domain, with only upper limits on the relative peak flux increase obtained at soft X rays (<200%; Bilous et al. 2012), soft (<250%; Lundgren et al. 1995), high (Bilous et al. 2011; <400%) and very high-energy gamma rays (<1000%; Aliu et al. 2012). From optical observations it seems that the observed energy excess in a GP is roughly the same in the radio as in the optical. If we scale from the optical to X-rays we consequently expect to see a similar flux increase. Our assumption is consistent with the first constraint on a GP strength in the Crab pulsar in the 15–75 keV energy range recently obtained with *Suzaku* in coincidence with a GRP (Mikami et al. 2014). The high counting rate from the Crab pulsar with the LAD and its 1 $\mu$s absolute timing accuracy will allow for individual pulses studies. Thus, it will be possible to detect, for the first time, GPs in X-rays from the Crab and in other relatively bright X-ray pulsars, monitor their correlation with GPs observed simultaneously in radio with the *SKA*, and possibly in the optical with the *E-ELT* (e.g., Mignani 2010), and determine the correlation between coherent and incoherent radiation production mechanisms in the neutron star magnetosphere. We ran a set of simulations to evaluate the capability of the LAD to detect GPs in the Crab pulsar light curve. We computed the number of counts in the 2.0–50.0 keV energy band in a 10 ks integration for both the pulsed emission and the nebula, from the count rate derived with the `fakeit` command in XSPEC, and the background. We produced a set of 300000 light curves with 100 phase bins randomising the number of counts in one period for all these three components. Figure 2 shows the average of the 300000 simulated light curves. From this set, we produced the expected distribution of pulsed counts in the bin centred on phase 0.0, which we define as coincident with the peak of the main radio pulse, deriving the off-pulse counts from the phase interval 0.6–0.8. Then, we reproduced a set of 10000 more light curves adding a fixed number of pulsed counts at phase 0.0, corresponding to a 3% peak intensity increase, with a GP frequency of 1 Hz. We finally averaged these 10000 light curves and compared the distributions at phase 0.0 with the previous one. As seen from Fig. 2, we can clearly distinguish the two distributions for a 3% peak flux increase at 13.7 $\sigma$, and are sensitive to a peak flux increase as small as ~1% at the 4.5$\sigma$ level. A similar analysis for the 1.6 ms pulsar PSR B1937+21 at the observed 0.13 Hz GRP rate (Soglasnov et al. 2004) indicates a $5\sigma$ detection for a 6% GRP enhancement.

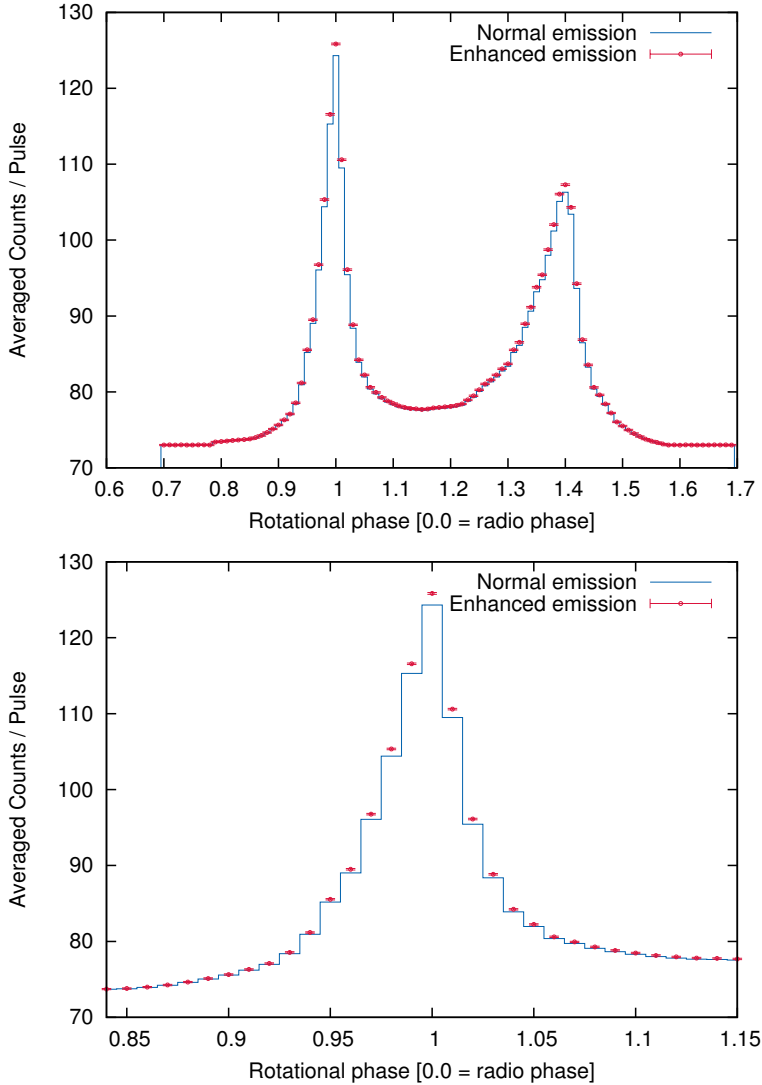

Figure 2: *Top:* Light curve of the Crab pulsar obtained from the average of 300000 randomly simulated LAD light curves (blue line) compared with the average of 10000 simulated LAD light curves (red circles) corresponding to a 3% peak count enhancement at phase 0.0, defined as the peak of the main radio pulse, due to the occurrence of a GP. The errors on the average GP counts are about the same size as the symbols, whereas the errors on the average normal emission is not recognisable on this scale *Bottom:* Same plot but centred on the phase of the main radio pulse.

No X-ray emission has been observed yet from nulling and intermittent radio pulsars (e.g., Rea et al. 2008; Lorimer et al. 2012), whereas both radio and X-ray mode changes have been very recently found in PSR B0943+10 (Hermsen et al. 2013; Mereghetti et al. 2013), resulting in a factor of two X-ray flux increase. However, with an expected LAD count rate of $\sim 0.06\,\mathrm{counts\,s^{-1}}$ (2–10 keV) for this pulsar, the detection of the flux variation would be limited by the systematic uncertainty on the LAD background ($\sim 0.6\,\mathrm{counts\,s^{-1}}$; at 1% level). A possible mode change has been found in the RQ gamma/X-ray pulsar PSR J2021+4026 (Abdo et al. 2009; Lin et al. 2013), with a $\sim$20% drop in the gamma-ray flux simultaneous to an increase in the spin period derivative (Allafort et al. 2013). In this case, owing to the lack of suitable X-ray observations before/after the event we have no constraint on the X-ray variability. With a simulated count rate of $\sim 0.24\,\mathrm{counts\,s^{-1}}$ (2–10 keV) for PSR J2021+4026, in 10 ks the LAD would detect flux variations of a factor of ten, larger than in gamma rays but not impossible a priori. Mode changes could also occur in brighter X-ray pulsars, though, for which no evidence of mode changes, either in radio or X-rays, has been found, or searched for, yet. For instance, for a pulsar with a 2–10 keV count rate of $\sim 2\,\mathrm{counts\,s^{-1}}$, the LAD would detect variations as low as a few tens per cent in 10 ks.

About 70 long-period (2–10 s) pulsars, dubbed RRaTs, have been detected thanks to the emission of sporadic, short radio bursts (< 1 s), rather than by their persistent pulsed radio emission. The physical mechanism driving the RRaT-like emission is unclear, although is becoming more and more established that RRaTs are the tip of the iceberg of the population of RPPs, and are possibly related with transient radio magnetars. While many more RRaTs will be discovered by *SKA*, only one has been detected as an X-ray pulsar so far: that with the highest magnetic field (McLaughlin et al. 2007). Searches for X-ray pulsations from more RRaTs with the LAD, and for the X-ray counterpart to the sporadic radio bursts, together with simultaneous *SKA* observations, will be crucial to shed light on the physical processes responsible for the origin of the radio bursts and their connection with the steady, or possibly bursting X-ray emission.

## 5 Bursts and outbursts from magnetar magnetospheres

Magnetars are X-ray bright, but usually RQ, INSs, with typically longer spin periods (2–12 s), higher spin-down rates, and stronger magnetic fields ($\sim 10^{13} \ldots 10^{-15}$ G) – inferred from the magnetic dipole model – with respect to most RPPs, hence the name (see, e.g., Rea & Esposito 2011 and references therein). Like RPPs, magnetars show pulsed X-ray emission (Fig. 1) but, unlike the vast majority of RPPs, they are also highly variable in the X-rays, with the flux growing up to orders of magnitude. Such variability occurs on different times scales, e.g., in short bursts (< 1 s), intermediate flares (1–40 s), giant flares (~500 s), and outbursts (several weeks to years). While giant flares are rare and have not been seen to repeat, both bursts and intermediate flares can recur on a few hour time scales.

The LAD will be able to follow-up on bursts and intermediate flares after their detection by the WFM (Sect. 6). This will be crucial to search for cyclotron absorption lines in the X-ray spectrum possibly observed at energies above 4 keV (e.g., Ibrahim et al. 2003), which brings direct measurements of the magnetic field during a burst. We simulated a LAD spectrum of a magnetar burst assuming two blackbody (BB) spectra with $kT_1 = 5.4$ keV and $kT_2 = 10.2$ keV, the same as observed in the bursts of the magnetar SGR 1900+14 (Israel et al. 2008), and an absorption line at an energy of 4.5 keV with equivalent width (EW) as small as 200 eV. By assuming an $N_H = 2.1 \times 10^{22}$ cm$^{-2}$, an observed 2–50 keV X-ray flux as small as $2.4 \times 10^{-7}$ erg cm$^{-2}$ s$^{-1}$ (10 Crab), and a 0.06 s burst duration, our simulation yields a net count rate of $(7.56 \pm 0.04) \times 10^5$ counts s$^{-1}$. The fit to the simulated spectrum (Fig. 3, left) gives $N_H = (2.2 \pm 0.2) \times 10^{22}$ cm$^{-2}$, $kT_1 = 5.3 \pm 0.2$ keV, $kT_2 = 13^{+11}_{-4}$ keV, and EW = 220 eV, with a line significance of $> 14\sigma$.

Outbursts are likely produced by deformations of the neutron star crust, which, in turn, produces a rearrangement of the external magnetic field (Rea & Esposito 2011). The LAD will study the magnetar light curve and spectral evolution (including cyclotron absorption features) over a broad energy range, from a few hours after the onset of the outburst to its decay back to the quiescent state, exploiting a better time and spectral resolution than both *NuSTAR* (Harrison et al. 2013) and *Astro-H* (Takahashi et al. 2014). In particular, LAD phase-resolved spectroscopy will allow one to track the variability of the magnetosphere configuration during the outburst decay following the variations in the PL spectral component, and search for line variability. For objects that reach very high surface temperatures ($kT \sim 1$ keV), phase-resolved spectroscopy will also allow one to map the region of the neutron star surface heated during the early phases of the outburst.

A phase-variable absorption line in the X-ray spectrum of the magnetar SGR 0418+5729 in outburst has been recently discovered with *XMM-Newton* and interpreted as a proton cyclotron feature produced by a magnetic loop with a field ranging from $2 \times 10^{14}$ G to more than $10^{15}$ G (Tiengo et al. 2013). The energy of such a feature varies very rapidly with phase and could be detected only by dissecting the source spectrum in about 50 phase intervals. While the parameters of the absorption line could be well constrained at energies below ~2 keV, where the effective area of *XMM-Newton* is largest, the poor energy resolution of the *RXTE* Proportional Counter Array (PCA) hampered a detailed phase-resolved spectroscopy of the line up to ~10 keV. Extending this analysis to more objects is of paramount importance for understanding the magnetic field structure in neutron stars and requires an X-ray instrument with large effective area and a sufficient temporal and spectral resolution. The characteristics of the *LOFT*/LAD perfectly match these requirements at energies above 2 keV, making it possible the most sensitive search for variable (both in phase and energy) proton cyclotron features in neutron stars with surface magnetic fields exceeding 4 × $10^{14}$ G. This is well illustrated by a simulation of the LAD spectrum (Fig. 3, right) of one of the 50 phase-resolved spectra (600 s integration each) extracted from the 30 ks *XMM-Newton* observation of SGR 0418+5729 (Tiengo et al. 2013). We computed the LAD spectrum from the best-fit *XMM-Newton* model: a black body ($kT$ ∼ 0.9 keV) plus a PL ($\Gamma$ = 2.8) and a cyclotron absorption line at the energy of 4.5 keV (500 eV EW). This absorption line was only marginally detected and poorly constrained (> 20% uncertainty on the line centroid) in the *XMM-Newton* spectrum relative to this specific phase interval (0.09–0.11; Tiengo et al. 2013), while a LAD integration of the same duration as the *XMM-Newton* one (600 s) would lead to a robust detection, with the line centroid constrained within 1%.

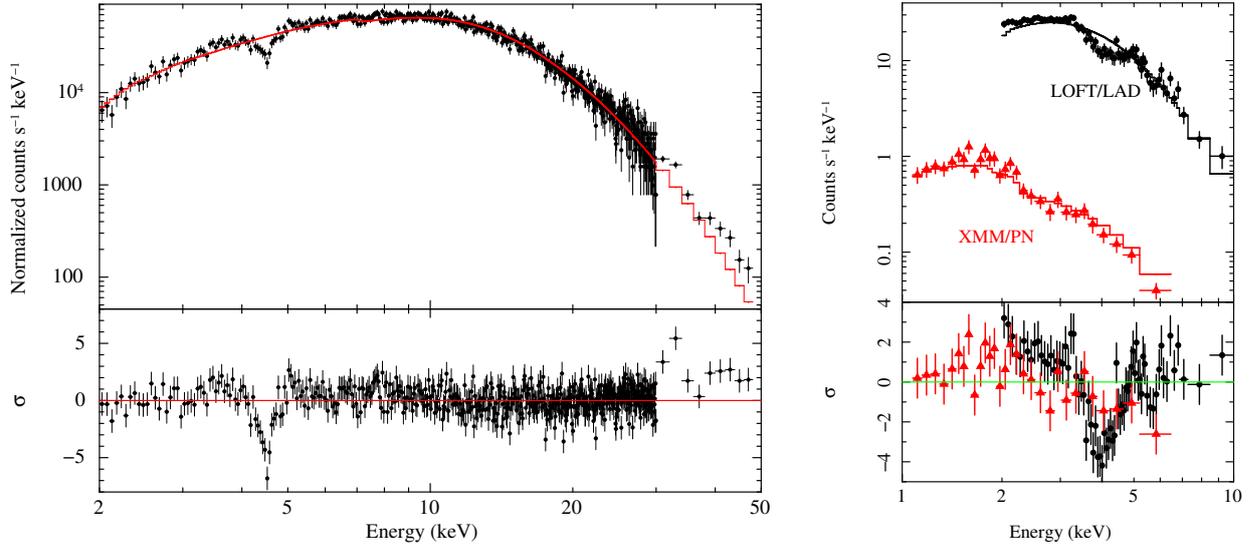

Figure 3: *Left:* Simulated LAD spectrum of a magnetar burst (black points) for a 0.06 s burst duration, assuming a flux of 10 Crab, a double black body spectrum ($kT_1$ = 5.4 keV and $kT_2$ = 10.2 keV) with $N_H$ = 2.1 × $10^{22}$ cm$^{-2}$, and an absorption feature at 4.5 keV (200 eV EW). The red line in the top panel shows the best-fit model. The lower panel shows the residuals with respect to the best-fit model. *Right:* Simulated LAD phase-resolved spectrum of SGR 0418+5729 in outburst (black points) in the phase interval 0.09–0.11 (600 s integration), assuming a black body ($kT$ = 0.9 kev) and a power law ($\Gamma$ = 2.8) with $N_H$ = 1 × $10^{22}$ cm$^{-2}$, and an absorption line at 4.5 keV of EW = 500 eV (Tiengo et al. 2013). The observed *XMM-Newton* EPIC-pn spectrum is shown by the red points. In both cases the solid lines indicates the model of the phase-averaged spectrum (Tiengo et al. 2013) rescaled to the flux of the selected phase interval. The lower panel shows the residual with respect to this model.

Monitoring the magnetar spin period is crucial to detect glitches, i.e., sudden changes in the spin period, and/or other irregularities in the star rotation and study their possible relation with the occurrence of short bursts and flares. Most magnetars have been discovered in the Galactic Plane. Thanks to the large field of view of the *LOFT*/WFM (4.1 sr), a large fraction of the Galactic Plane will be covered during most *LOFT* pointings. This gives the possibility of repeatedly measuring the spin period of several magnetars at the time and obtaining phase-connected timing solutions. We note that all the persistently bright magnetars are RQ and timing measurements can only be obtained in the X-ray band. Since these sources have typical X-ray fluxes of a few mCrabs, X-ray pulsations can be accurately measured by the WFM in few tens of ks. Since magnetars have spin periods (2–12 s) longer than RPPs, the full time resolution of the WFM (10 µs) is not strictly required, which relaxes the constraints on the usage of the on-board telemetry.

## 6 The search for unidentified isolated neutron stars

In the last 30 years, X-ray observations have been fundamental in discovering entirely new types of INSs (Harding 2013), whose existence would have passed unnoticed otherwise. Their study paved the way to novel studies in nuclear physics, electromagnetism, and massive stellar evolution and opened new fields, such as neutron star astroseismology, which is a pivot of one of the two *LOFT* Core Science Programmes (The neutron star equation of state). Therefore, searching for INSs among high-energy sources discovered by the present and future generation of satellites is crucial to further expand the horizons of neutron star astronomy.

RPPs are the most numerous class of known Galactic gamma-ray sources, and many more are expected to lurk among the thousands of sources still unidentified at the end of the *Fermi* mission (2018 or later). A large fraction of these unidentified gamma-ray pulsars is expected to be RQ. While some of them might actually be RL, simply too faint for current radio telescopes and eventually detected by the *SKA*, others might be genuinely RQ, with the radio beam pointing away from Earth and missed by radio pulsar surveys. A more realistic assessment of the size of the RQ gamma-ray pulsar population will shed light on the still mysterious nature of the unidentified Galactic gamma-ray sources and influence neutron star population synthesis models based on the expected Galactic supernova rates, so far mainly verified against radio pulsar surveys. The quest for these unidentified gamma-ray pulsars is an enduring task that will continue well after the end of the *Fermi* mission. For faint gamma-ray pulsars a direct identification through the detection of an unknown spin period is hampered by the low number of gamma-ray photons and position uncertainty (a few arcminutes), which affects the reconstruction of the time of arrival of the gamma-ray photons with respect to the Solar System Barycentre. The case is even more complex when searching for binary MSPs in unidentified *Fermi* sources since one has to account for the effects of the unknown orbital parameters on the time of arrival of the gamma-ray photons. Although all binary MSPs detected so far are RL, in some the radio emission might be quenched by plasma trapped in the tight binary system environment and not be easily detectable even in deep radio observations. This was the case for PSR J1311−34, whose detection as a gamma-ray pulsar (Pletsch et al. 2012) preceded that as a radio pulsar (Ray et al. 2013). Such problems can be tackled by looping "blind" gamma-ray periodicity searches over a uniformly spanned grid of trial positions, making a massive use of super-computing power facilities (Pletsch & Clark 2014). With *LOFT*, the best candidate gamma-ray pulsars, could be scanned for X-ray pulsations through single pointings with the LAD, whose 1° field of view easily covers the gamma-ray source error boxes, possibly taking advantage of reference X-ray source positions from the *eRosita* all-sky survey (Predehl et al. 2014). The detection of X-ray pulsations will then provide reference X-ray ephemeris to fold the gamma-ray data, yielding the discovery of pulsations also in the gamma rays. We simulated the detectability of X-ray pulsations from a LAD source in a blind search analysis. As done in Sect. 4, we assumed a single-peak light curve with a Lorentzian profile, variable PF and peak FWHM, a pulsar PL X-ray spectrum with $\Gamma_X = 1.5$. We ran a blind search for pulsations with periods in the 0.01–10 s range ($4 \times 10^7$ trial periods) through a Rayleigh test. For instance, for a 0.5–10 keV unabsorbed flux of $3 \times 10^{-13}$ erg cm$^{-2}$ s$^{-1}$, 45% PF and a 0.1 FWHM, we found that pulsations can be detected at $> 5\sigma$ in 10 ks, after accounting for the number of trials. Thus, the LAD will detect X-ray pulsations for about 30 of the best candidate gamma-ray pulsars selected on the basis of logistic regression techniques (Salvetti et al., in preparation), assuming the X-ray brightness estimated by scaling the X-ray to gamma-ray flux ratio of known gamma-ray pulsars (Abdo et al. 2013).

With a similar sensitivity, the LAD will search for X-ray pulsars in several Galactic and Magellanic Cloud sources detected by the Cherenkov Telescope Array (*CTA*; Acharya et al. 2013), which will become operational in 2018. In the very high-energy (VHE) domain, the so-called Pulsar Wind Nebulae (PWNe) are the second most numerous class – 37 among the 154 sources currently detected with Cherenkov telescopes (http://tevcat.uchicago.edu/). Consequently, PWNe are expected to be a large fraction of the new sources anticipated in Galactic plane surveys with *CTA*. Thanks to its energy range (20 GeV to hundreds of TeV), angular resolution (up to 0.02°), full-sky survey $5\sigma$ sensitivity (~0.01 Crab), the *CTA* will increase the population of the Galactic

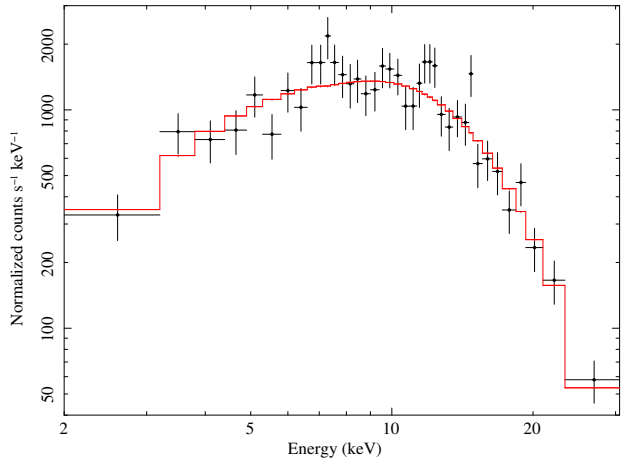

Figure 4: Simulated WFM spectrum of a magnetar burst (black points), assuming a burst duration of 0.6 s, a 2–50 keV flux of 150 Crab, and a two-BB spectrum ($kT_1$ = 5.4 keV, $kT_2$ = 10.2 keV) with $N_H$ = 2.1 × $10^{22}$ cm$^{-2}$. The solid red line represents the best-fit to the data.

PWNe detected at VHE by about an order of magnitude. The estimate of the number of PWNe detectable with a $5\sigma$ significance with *CTA*, based upon an assumed model of their Galactic distribution and their assumed properties, yields from 300 to 600 PWNe (see de Ona-Wilhelmi et al. 2013). This unprecedented database will be key to constrain the amount of energy carried by the pulsar relativistic wind for different types of pulsars and study the mechanisms of production of VHE radiation from the wind interaction with different types of environments. The identification of a *CTA* source as a PWN, however, requires the detection of its associated pulsar, either at GeV, radio, or X-ray energies. PWN candidates detected at VHE by *CTA* will be, thus, obvious targets to search for X-ray pulsations with the LAD, which would unveil the presence of a pulsar and certify the identification of the *CTA* source as a PWN.

The LAD will also point at several of the most interesting targets selected in the sample of about 10000 young radio pulsars and 1500 MSPs that will be discovered during the Phase I of *SKA*, due to be completed by the time *LOFT* will be launched. The aforementioned pulsar samples will then be roughly doubled during Phase II of *SKA*, which is expected to run at the same time as the *LOFT* mission, enlarging the current INS population by about one order of magnitude (Tauris et al. 2014; Keane et al. 2014). With these very large samples, *SKA* will finally enable to tackle long-standing and still unsolved issues in RPP emission theory, like the geometry of the radio beam(s), the connection between the radio emission processes and the variability in the pulsar emission (i.e., the erratic emission behaviour described in Sect. 4), the link (if any) of a pulsar magnetosphere status and its timing properties, as well as if (and to which extent) the properties of the magnetosphere change from one class of pulsars to another (Karastergiou et al. 2014). With the same sensitivity as described in Sect. 3, the LAD will follow up a significant number among the closest and brightest of the RPPs detected by the *SKA*, finally leading to build self-consistent models of the neutron star magnetosphere, along the rationale anticipated in Sections 3 and 4.

After the discovery of the prototype source (Mazets et al. 1979), about 30 magnetars (http://www.physics.mcgill.ca/~pulsar/magnetar/main.html) have been found, including candidates (Olausen & Kaspi 2014). The unexpected discovery of both radio-loud (Camilo et al. 2006) and low-magnetic field magnetars (Rea et al. 2010) has triggered a profound rethinking of the nature of these objects and the path to their understanding still passes through the discovery of more sources, unveiling similarities and diversities. Most magnetars have been discovered thanks to their transient X-ray emission, and previous large field-of-view telescopes, such as the All-Sky Monitor aboard *RXTE* or the Burst and Alert Monitor on *Swift*, were instrumental in this role. Thus, we expect that several "dormant" magnetars are yet to be discovered. The WFM is expected to discover at least a new candidate magnetar per year, triggering alerts for other new facilities. We simulated a 0.6 s-long magnetar burst as observed by the WFM (Fig. 4). We assumed a 2–50 keV X-ray flux of 150 Crab and, as in Sect. 5, a two BB spectrum ($kT_1$ = 5.4 keV, $kT_2$ = 10.2 keV), with $N_H$ = 2.1 × $10^{22}$ cm$^{-2}$. The input parameters can be only partially recovered by the fit to the simulated spectrum (e.g., the best-fit flux is $130 \pm 15$ Crab), but its very high signal-to-noise ratio, with a net count rate $(0.95 \pm 0.05) \times 10^4$ counts s$^{-1}$, clearly shows that the WFM will be able to detect magnetar bursts with high significance within the required short integration times.